\documentclass[prl,aps,twocolumn,10pt,showpacs,superscriptaddress,amsmath,amssymb]{revtex4-1}
\usepackage{amsmath}
\usepackage{graphicx,amssymb}
\usepackage[pdftex]{color}
\usepackage{color}
\begin{document}
\bibliographystyle{apsrev}

\title{Charge transfer induced interfacial ferromagnetism in La$_{0.7}$Sr$_{0.3}$MnO$_3$/NdNiO$_3$ }

\author{K.~Chen}
\email{kaichen.hzg@gmail.com}
\affiliation{Helmholtz-Zentrum Berlin f$\rm \ddot{u}$r Materialien und Energie, Albert-Einstein-Strasse 15, 12489, Berlin, Germany}
\author{C.~Luo}
\affiliation{Helmholtz-Zentrum Berlin f$\rm \ddot{u}$r Materialien und Energie, Albert-Einstein-Strasse 15, 12489, Berlin, Germany}
\author{B.~B.~Chen}
\affiliation{MESA$^{+}$ institute for nanotechnology, University of Twente, PObox 217, Enschede, The Netherlands}
\author{R. M.~Abrudan}
\affiliation{Helmholtz-Zentrum Berlin f$\rm \ddot{u}$r Materialien und Energie, Albert-Einstein-Strasse 15, 12489, Berlin, Germany}
\author{G.~Koster}
\affiliation{MESA$^{+}$ institute for nanotechnology, University of Twente, PObox 217, Enschede, The Netherlands}
\author{S.~K.~Mishra}
\affiliation{School of Material Science and Technology, Indian Institute of Technology (BHU), Varanasi 221005, India}
\author{F.~Radu}
\email{florin.radu@helmholtz-berlin.de}
\affiliation{Helmholtz-Zentrum Berlin f$\rm \ddot{u}$r Materialien und Energie, Albert-Einstein-Strasse 15, 12489, Berlin, Germany}

\date{\today}

\begin{abstract}

Charge transfer induced interfacial ferromagnetism and its impact on the exchange bias effect in La$_{0.7}$Sr$_{0.3}$MnO$_3$/NdNiO$_3$ correlated oxide heterostructures were investigated by soft x-ray absorption and x-ray magnetic circular dichroism spectra in a temperature range from 10 to 300~K. Besides the antiferromagnetic Ni$^{3+}$ cations which are naturally part of the NdNiO$_3$ layer, Ni$^{2+}$ ions are formed at the interface due to a charge transfer mechanism involving the Mn element of the adjacent layer. They exhibit a ferromagnetic behavior due to the exchange coupling to the Mn$^{4+}$ ions in the La$_{0.7}$Sr$_{0.3}$MnO$_3$ layer. This can be seen as detrimental to the strength of the unidirectional anisotropy since  a significant part of the interface does not contribute to the pinning of the ferromagnetic layer. By analyzing the line shape changes of the x-ray absorption at the Ni L$_{2,3}$ edges, the metal-insulator transition of the NdNiO$_3$ layer is resolved in an element specific manner. This phase transition is initiated at about 120~K, way above the paramagnetic to antiferromagnetic  transition of NdNiO$_3$ layer which  measured to be 50~K. Exchange bias and enhanced coercive fields were observed after field cooling the sample through the N{\'e}el temperature of the NdNiO$_3$ layer. Different from La$_{0.7}$Sr$_{0.3}$MnO$_3$/LaNiO$_3$, the exchange bias observed in La$_{0.7}$Sr$_{0.3}$MnO$_3$/NdNiO$_3$ is due to the antiferromagnetism of NdNiO$_3$ and the frustration at the interface. These results suggest that reducing the interfacial orbital  hybridization may be used as a tunable paramater for the strength of the exchange bias effect in all-oxide heterostructures which exhibit a charge transfer mechanism.   

\end{abstract}
\maketitle

\section{introduction}
Electronic reconstruction at oxide interfaces~\cite{Mannhart2010}, has been intensively investigated to explain the appearance of novel properties in layered structures, such as the ferromagnetic metallic state at the interface of LaAlO$_3$/SrTiO$_3$~\cite{Brinkman2007, Basletic2008, Bert2011,Li2011}, and the interfacial two-dimensional electron gas in LaAlO$_3$/SrTiO$_3$\cite{Ohtomo2004} and $\gamma$-Al$_2$O$_3$/SrTiO$_3$~\cite{Chen2013, Lee2019} systems. Oxygen vacancies \cite{Sirena2009, Guo2016, Yao2017}, structural strains \cite{Peng2015}, and charge redistributions \cite{Pentcheva2008} in the atomic layers close to the interfaces have been established as underlying mechanisms for the complex traits of oxide-based heterostructures. A sharp bandwidth-controlled metal-insulator transition (MIT)~\cite{Imada1998} has been widely observed in the RNiO$_3$ (RNO) nickelate oxides (both in bulk and films) with smaller lanthanide ions (R$\neq$La), accompanied by a transition from paramagnetic to antiferromagnetic (AFM) state as a function of temperature. Charge transfer and orbital reconstruction mechanisms that occur at the interface serves as the basic mechanisms for   numerous fascinating phenomena that have been observed in ferromagnetic (FM)/RNiO$_3$ heterostructures, like metal-insulator-metal transition in CoFe$_2$O$_4$/NdNiO$_3$ \cite{Saleem2017}, exchange bias effect~\cite{Sanchez2012, Zhou2017}, noncollinear magnetic structure \cite{Hoffman2016}, and superconductivity \cite{Zhou2018} in La$_{1-x}$Sr$_{x}$MnO$_3$/LaNiO$_3$~(LSMO/LNO). Moreover, the charge transfer at the La$_{0.8}$Sr$_{0.2}$MnO$_3$/NdNiO$_3$ interface can be controlled via strain engineering~\cite{Xu2018}. 

For the charge transfer scenario of the  La$_{0.7}$Sr$_{0.3}$MnO$_3$/NdNiO$_3$ (LSMO/NNO) bilayer, hole transfer from Ni$^{3+}$ to Mn$^{3+}$ would result in Ni$^{2+}$ and Mn$^{4+}$ at the interface favoring the ferromagnetic Ni$^{2+}$-O$^{2-}$-Mn$^{4+}$ interactions, similar to that observed in La$_2$NiMnO$_6$ \cite{Dass2003, Das2008}. Such an interfacial ferromagnetism of Ni$^{2+}$ has been observed in (LaNiO$_3$)$_n$/(LaMnO$_3$)$_2$ superlattices \cite{Hoffman2013}. The FM interaction between Ni$^{2+}$ and Mn$^{4+}$ which are further coupled to the the antiferromagnetic  Ni$^{3+}$ ions in the buried NdNiO$_3$ layer, give rise to frustrated magnetic regions that affect the coercive and the exchange bias field of the FM layer \cite{rz:2008,Ning2015, Peng2015}. 
An exchange bias effect is conventionally observed for a ferromagnetic layer which shares a common interface with a antiferromagnetic one. As such, the NNO layer can be used as the antiferromagnetic bias layer which mediates an eventual occurrence interfacial unidirectional anisotropy. By contrast, LNO is a  Pauli paramagnetic, therefore it is not expected that this magnetic ground state will support the formation of an unidirectional anisotropy. Nevertheless,  by surprise an exchange bias was observed in LSMO/LNO~\cite{Sanchez2012, Zhou2017,Xu2018}, and it is yet  unclear what is the underlying mechanism and if it offers sufficient tunability of the unidirectional magnetic anisotropy. By contrast, the NNO thin films does exhibit an antiferromagnetic ground state which may be optimized against strain, dimensionality, and other intrinsic and extrinsic constrains \cite{Kumar2010, Liu2010, Liu2013, Dhaka2015, Wang2015, Heo2016, Hooda2016, Palina2017, Onozuka2019, Li2019}.

In this paper, we report on the complexity of the magnetic interactions that occur at the interface, revealing correlations between the unidirectional anisotropy, charge transfer, interfacial exchange coupling, and the metal insulator transition in LSMO/NNO heterostructures. In the next section we introduce  the samples  and describe their structural, magnetic and transport  properties measured by means of laboratory tools. In the third section we make use of soft x-ray spectroscopy to reveal an interfacial charge transfer between the constituent  Mn the Ni elements and  demonstrate the occurrence of the MIT phase transition for the NNO layer. By analysing the XAS at the Mn edge for two different thickness of the LSMO layer, we bring compelling evidence for a valence gradient of the Mn element towards the interface. Moreover, by analysing the peak positions in the XAS spectra measured across the Ni L$_{2,3}$ resonant edges we observe the formation of  Ni$^{2+}$.  Corroborating these two observations we are able to fully demonstrate the occurrence of a charge transfer effect at the LSMO/NNO interface. Furthermore, by analysing the profile of the XAS spectra which were measured as a function of temperature, we clearly reveal the occurrence of the MIT phase transition which initiates at 120~K. In the fourth section we involve x-ray magnetic circular dichroism measured at the Mn, Ni and Nd edges measured as a function of temperature, and as a function of an external magnetic field reaching up to 8~Tesla. XMCD measured for the Mn layer as a function of temperature provides the Curie temperature of the FM layer which was 180~K. Through the XMCD measurements as a fuction of an external field and as a function of temperature at the Nd Ni M$_{5,4}$ resonant edges, we are able to measure the N{\'e}el temperature of the AF NNO layer which is equal to 50~K for our system. The  measurement of the three critical temperatures within the same methodical environment for the same sample is an important ingredient for disentangling key contributions to the magnetic interactions at the interface of this oxide heterostructure. For instance, we will show that ferromagnetic Ni$^{2+}$-O$^{2-}$-Mn$^{4+}$ interaction at the interface is observed below 180~K, and before the onset of the AFM order in the NNO layer. Also, the exchange bias effect is observed below the para-antiferromagnetic transition of NNO at T$_N$=50 K, after field cooling the sample in an external magnetic field.  Thus, we will be able to conclude that the exchange bias is related to the onset of an AFM ordering of NNO, governed by the exchange coupling between the constituent antiferromagnetic Ni$^{3+}$ and Nd ions,  and the ferromagnetic LSMO layer that shares a common interface with the NNO layer. The interfacial magnetic frustration will be inferred from the occurrence of mixed ferromagnetic and antiferromagnetic ordering, supporting also possible spin glass states formed at the LSMO/LNO \cite{Ning2015, Peng2015}, LaMnO3/LNO~\cite{Lee2013, Gibert2016, Zang2017}, and LSMO/SrMnO$_3$ interface~\cite{Ding2013}. 

\begin{figure}[b]
\includegraphics[width=1\linewidth]{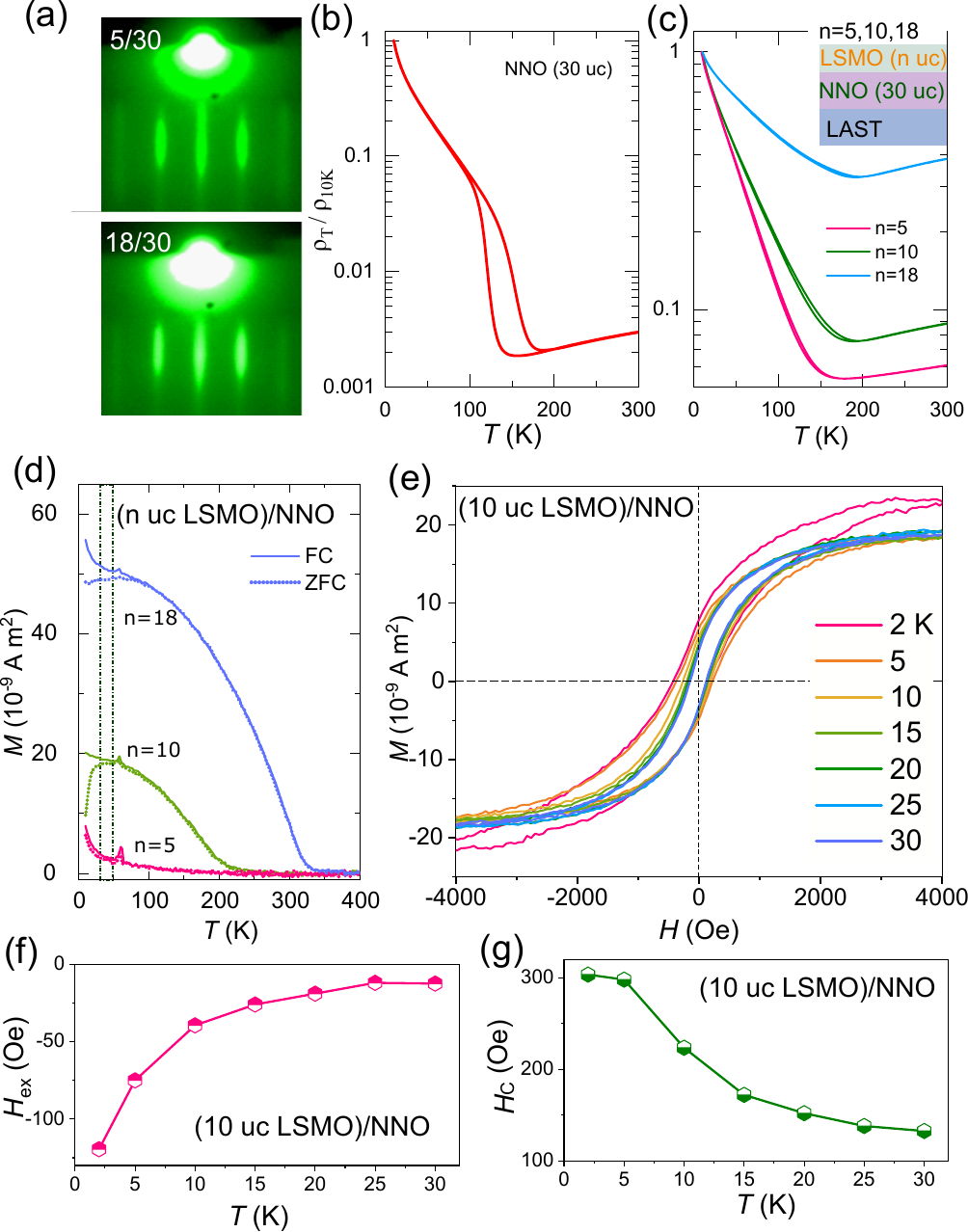}
\caption{\small{(a) In-situ RHEED characterization of the oxide heterostructures for (5 uc LSMO)/NNO and (18 uc LSMO)/NNO grown on LSAT (001) substrates and the T-dependent resistance measurement as the evidence of Mott transition in bare NNO (b) and LSMO/NNO bilayers (c).  (d) M-T curve for LSMO/NNO bilayers for ZFC/FC conditions. The dashed area highlighting the modification in magnetism below 50K, and (e) the M-H hysteresis loops for (10 uc LSMO)/NNO at various temperatures after the FC procedure, with the variation in exchange bias (H$_{ex}$) (up to 120 Oe at 2 K)and coercivity field (H$_C$) as a function of temperature shown in (f) and (g), respectively.}}
\label{fig:1}
\end{figure}

\section{Samples description and Exchange bias effect in LSMO/NNO}

LSMO/NNO bilayers have been prepared by pulsed laser deposition system at MESA$^{+}$, University of Twente, The Netherlands using an ultra-high vacuum chamber. NNO films with constant thickness equal to 30 unit cell (uc) were deposited onto  $\rm (LaAlO_3)_{0.3}(Sr_2TaAlO_6)_{0.7}$ (LSAT) (001) substrates. The ferromagnetic LMSO layer with thickness of 5, 10 and 18 uc were deposited on top, therefore sharing a common interface with the antiferromagnetic NNO layer. Both constituent films were grown at 700 $^{\circ}$C, in an oxygen pressure of 0.2 mbar and for a laser fluence which was set to $\sim$2 J/cm$^2$. 

Structural characterization of the oxide bilayers has been carried out using reflection high-energy electron diffraction (RHEED) (Fig.~1a), X-ray diffraction and  atomic force microscopy(not shown), suggesting that a high structural quality has been achieved for all the  oxide heterostructures. Fig.~1b and ~1c show the electrical transport measurements of the 30 uc NNO reference layer and of the oxide bilayer samples, which were performed in a  Van der Pauw geometry using a constant current source. Transport measurements demonstrate the occurrence of the metal-insulator transition (MIT) for all bilayer samples, showing that  the transition temperature  takes place at about 120~K. Notice that the pronounced irreversible nature of MIT for bare NNO, seen as a broad thermal hysteresis of the resistivity measurements is strongly diminished for all investigated oxide LSMO/NNO bilayers. 

The magnetization as a function of temperature after zero field cooling (ZFC) and field cooling (FC) procedures has been recorded using a Quantum Design Superconducting Quantum Interference Device (SQUID) magnetometer. The field cooling and measuring field were both set to 500 Oe and the results are shown in Fig.~1d. The Curie temperatures T$_c$ are about 70, 180 and 290~K for (n uc LSMO)/NNO with n=5, 10, and 18, respectively. Besides the ferromagnetic behavior which originates from the top LSMO layers, a difference between ZFC magnetization and FC magnetization curves can be observed below 50~K, for all oxide bilayer samples. Using Vibrating Sample Magnetometer (VSM), magnetic hysteresis loops for the (10 uc LSMO)/NNO sample have been  measured from 2~K to 30~K after field cooling the sample in a an external field of 5000~Oe along [100] direction, as shown in Fig.~1e. An exchange bias field (H$_{ex}$) as well as an enhanced coercive field (H$_C$) were observed and they are both increasing as the temperature decreases,  as shown in Figs.~1f~and~1g, respectively. Note that the exhange bias field cease to exists at about 30~K, which is lower as compared to the critical temperature where the ZFC and FC curves deviates from each other (50~K). It can be assumed that the 50~K corresponds to the N{\'e}el temperature of the NNO and that 30~K corresponds to the blocking temperature for the exchange bias effect~\cite{rz:2008}. This will be confirmed later in the manuscript by measuring the N{\'e}el temperature of the NNO within an element specific susceptibility approach.

\section{Charge transfer and Metal-insulator transition}

\begin{figure*}[t]
\includegraphics[width=0.95\linewidth]{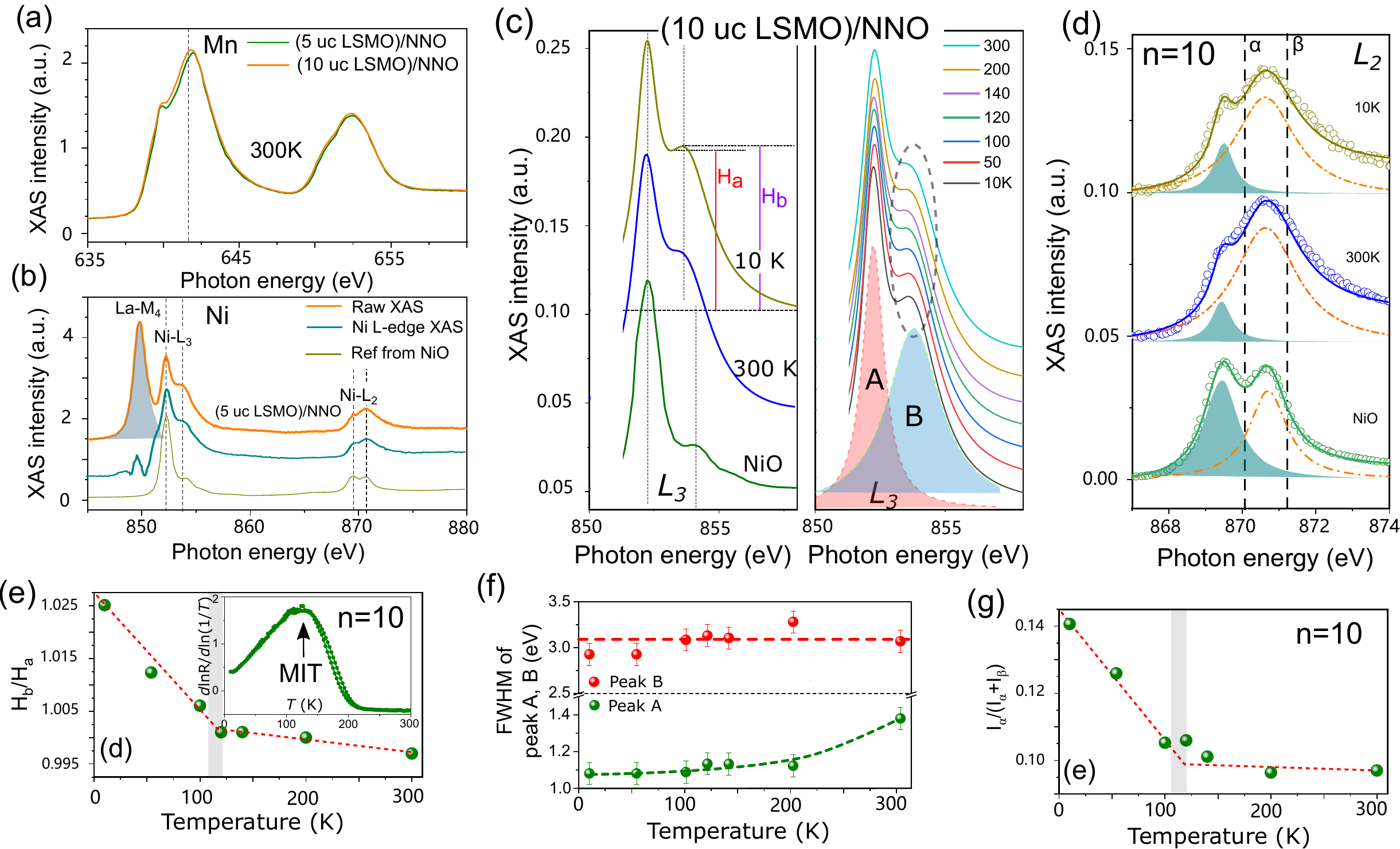}
\caption{\small{XAS measured in total electron yield (TEY) mode of Mn (a) and Ni (b) ions, with an observable difference between the Mn XAS for (5 uc LSMO)/NNO and (10 uc LSMO)/NNO. After removing the La-M$_4$ white lines from Ni XAS, the main contribution with Ni$^{2+}$ 3d$^7$ configuration at the interface is confirmed when comparing to the spectra from NiO. (c) Temperature dependent L$_3$-XAS spectra of Ni cations, with a well separated shoulder peak at the energy of 853~eV at lower temperatures. (e) The temperature dependence of the value of H$_b$/H$_a$ follows the MIT from the [$d$lnR/$d$ln(1/$T$) vs $T$] curve (inset), with H$_a$ and H$_b$ represent for the height of the valley and peak for the shoulder structure marked in (c), and (f) the temperature dependent of the width for peak A and B. (d) L$_2$-XAS spectra of the Ni cations at 10 and 300K and NiO, can be well reproduced from two separated peaks ($\alpha, \beta$) with the Lorentzian shape and (g) the temperature dependence of the value for I$_{\alpha}$/(I$_{\alpha}$+I$_{\beta}$), with I$_{\alpha}$ and I$_{\beta}$ represent the peak area of features $\alpha$ and $\beta$, respectively.} }
 \label{fig:2}
\end{figure*}

To shed light on the interfacial magnetic contributions to the exchange bias and enhanced coercive field observed for the oxide bilayer, we performed  XAS as well as the XMCD measurements at the 3d transition metal (Ni, Mn) L$_{2,3}$ edges and at the M$_{4,5}$ edges of Nd, as a function of temperature and for  external magnetic fields up to 8T. 
XAS were measured in total electron yield (TEY) mode with the photon polarization parallel (E//ab) to the samples to determine the valence states of Mn and Ni ions. The dichroic XMCD signal was recorded as the difference of the x-ray absorption spectra measured under a magnetic field of $\mu_0$H=8T applied perpendicular to the sample surface and with a parallel ($\sigma^+$) or antiparallel ($\sigma^-$) circular helicity beam. Note that the bare XAS spectra which serve as a probe for the MIT were measured with linear polarized beams. The XAS and XMCD spectra were recorded at the VEKMAG end station installed at the PM2 beamline, BESSY II, HZB~\cite{rn:2017}.

The information on the unoccupied Mn and Ni 3d states as well as the related valence states can be deduced from their L-edge absorption spectra, as shown in Fig.~2a and 2b, respectively. For Mn, the spectra correspond to on-site transitions from 2p$^6$3d$^n$ to 2p$^5$3d$^{n+1}$ and show two groups of multiplets, namely the L$_3$ (641-645 eV) and L$_2$ (652-656 eV) white line regions, split by the spin-orbit interaction of the Mn 2p core level. We find a significant difference between the Mn XAS (at 300K) for (5 uc LSMO)/(NNO) (X-curve) and (10 uc LSMO)/NNO (Y-curve) (normalized by the edge jump) with the former one being shifted towards higher energy values. This is indicative for different valence states carried by the Mn ions in these samples. Since the TEY mode is more sensitive to the surface, more interfacial information contributes to the spectra of (5 uc LSMO)/NNO sample. The shift of 0.6 eV for sample (5 uc LSMO)/NNO suggests for a higher contribution of Mn$^{4+}$ at the interface. This result clearly indicates that a valence gradient of Mn ions in LSMO occurs, with a higher valence at the interface and a lower valence in the outermost monolayers.

The XAS of Ni L-edge at 10K for NNO and the refence sample of NiO are shown in Fig.~2b. These spectra contain also the La-M$_4$ white lines located at 850.6 eV which have been removed for the consequent analysis. The XAS spectra at the L$_{2,3}$ edge of the transition metal oxides are highly sensitive to the valence state with an expected energy shift more than 1 eV~\cite{Guo2018} between the spectra of Ni$^{2+}$ of NiO and Ni$^{3+}$ of NdNiO$_3$. Here, the Ni L$_3$ main peak exhibits the same position as NiO with an octahedral-coordination oxygen environment, suggesting that the main contribution corresponds to Ni$^{2+}$ (3d$^{7}$) configuration. The Ni$^{3+}$ contribution from the deeper NNO monolayers is also visible as a shoulder at a higher energy equal to 854~eV. Two separated peaks can also be well observed at Ni L$_2$-edge,  at the same energy of Ni$^{2+}$. These two peaks may originate from the interfacial Ni$^{2+}$ or from  deeper NNO monolayers with insulating Ni$^{3+}$ states. A temperature dependent XAS investigation across the MIT will help to to resolve the origin of the two possible contributions to the double peak at Ni L$_2$ edge. Since the leading shoulder peak will disappear for the metallic Ni$^{3+}$ states, a variation of their relative weights will provide a self-consistent separation of those components. At this stage, corroborating the higher Mn valence at the interface and the formation of Ni$^{2+}$, we are able to confirm that a charge transfer occurs from Mn cations to Ni cations at the LSMO/NNO interface, fulfilling the charge balance requirement.

Although the main contribution of the Ni L$_{2,3}$-edge XAS is from the Ni$^{2+}$ formed at the interface of LSMO/NNO, one can still extract the information about the MIT of the buried NNO monolayers  from the change of the XAS line shape as a function of temperature. The temperature dependent spectra of Ni L$_3$-edge XAS (after removing the La M$_4$-edge) are shown in Fig.~2c. When compared to the spectra recorded at 300 K, the spectra measured at 10 K show a well separated shoulder peak at the energy of 853~eV. This type of peak splitting is observed throughout the nickelate series and has been associated with the charge-transfer energy separating the O~2p and Ni 3d states near the Fermi level \cite{Freeland2016}. In Fig.~2c, H$_a$ and H$_b$ are marked by lines, representing the height of the valley and the peak height for the shoulder structure. The temperature dependence of the ratio H$_b$/H$_a$ is shown in Fig.~2e. Upon cooling, it exhibits a rapid increasing below 120K, in agreement with the MIT temperature obtained from the [$d$lnR/$d$ln(1/$T$) vs $T$] curve shown in the inset. This cab be  due to an increase of the gap between the valence and conduction electrons bands or due to the delocalization of the two bands (3d$^7$ and 3d$^8\underline{L}$) when entering into the insulator phase, which leads to well separated peaks and to an enhanced peak to valley ratio H$_b$/H$_a$. The L$_3$-edge XAS can be well fitted with two Lorentzian shaped components, marked as peak A and B in Fig.2c . We observe that  the full width at half maximum (FWHM) of the Peak A does significantly change as a function of temperature, whereas the FWHM of the Peak B remains rather constant within the analysis accuracy. Also, the energy difference between the peak positions is not changing from 300 to 10K (not shown).  These results suggests that through the MIT of the bulk NNO, the reduced mixing of d$^7$ and d$^8\underline{L}$ at the interface is due to the delocalization of the first component only, which is also consistent with the analysis at the L$_2$ edge which is described further below.

Similar to the L$_2$-edge XAS of NiO, two well separated Lorentzian shaped peaks ($\alpha, \beta$)  can well reproduce the XAS spectra of the Ni cations at the interface, as shown in Fig.~2d. Note that both Ni$^{2+}$ and Ni$^{3+}$ may contribute to these two peaks of $\alpha$ and $\beta$~\cite{Liu2011}. Since no valence change has been observed for Mn sites between 10 and 300K, we assume that the valence of Ni sites at the interface keeps unchanged through the MIT. Considering that the contribution to the $\alpha$ peak by Ni$^{2+}$ remains constant as a function of the temperature, than the temperature variation of the $\alpha$ peak intensity is due to the MIT of the Ni$^{3+}$ in the buried NNO monolayers. This scenario is also supported by  the temperature dependence of the  Peak A and B widths at the L3 edge (see Fig.~2f). The relative increase of the  $\alpha$ peak intensity at lower temperature can be observed according to the value of I$_{\alpha}$/(I$_{\alpha}$+I$_{\beta}$), where I$_{\alpha}$ and I$_{\beta}$ represent the peak area of $\alpha$ and $\beta$ peaks, respectively. The change of I$_{\alpha}$/(I$_{\alpha}$+I$_{\beta}$) as a function of temperature is shown in Fig.~2g. Its variation as a function of temperature  agrees with the character of H$_b$/H$_a$ shown in Fig.~2e, revealing that the same MIT occurs for the deeper NNO monolayers.

\begin{figure}[b]
\includegraphics[width=0.92\linewidth]{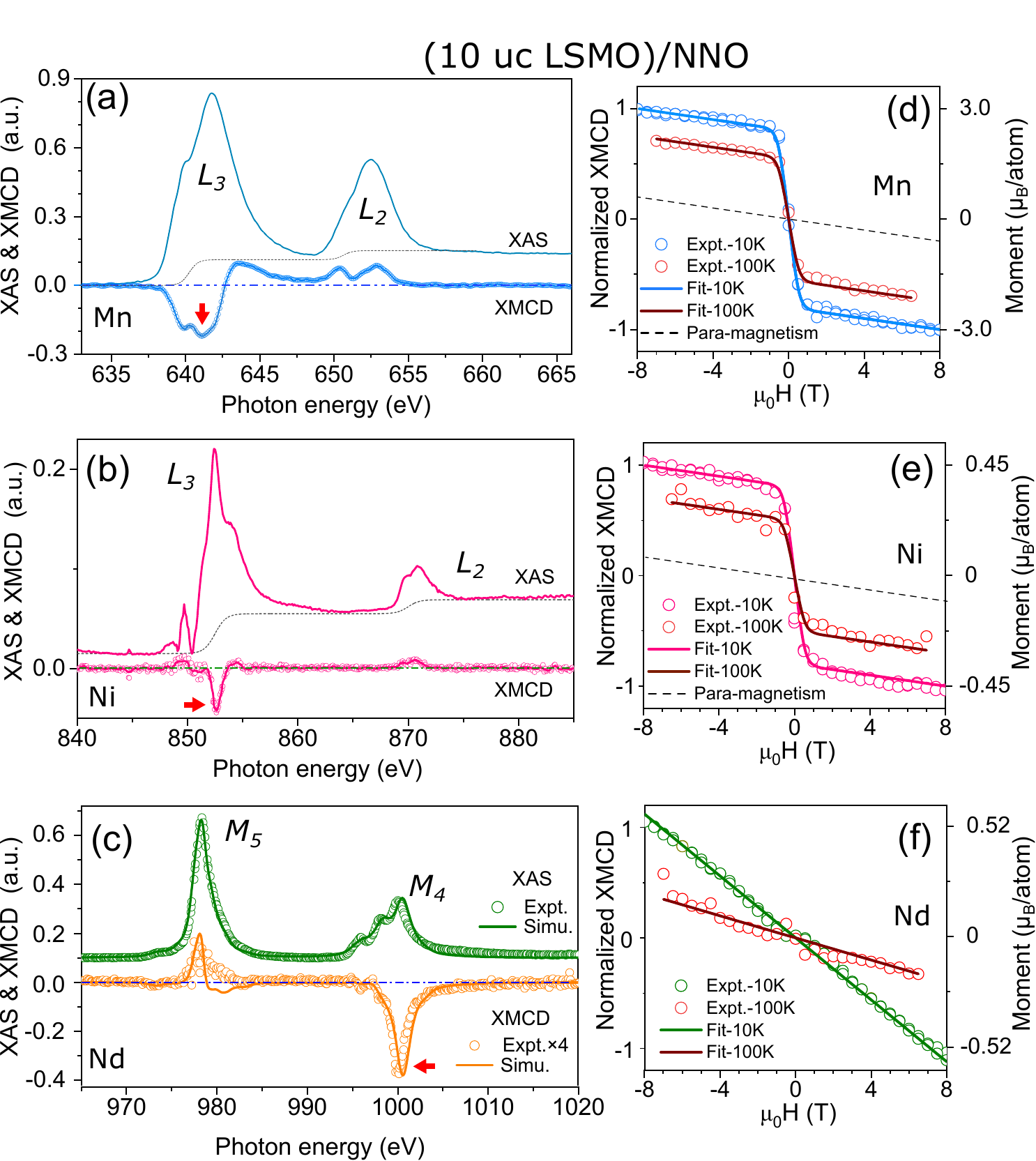}
\caption{\small{(a-c) XAS spectra of Mn L$_{2,3}$, Ni L$_{2,3}$, and Nd M$_{4,5}$-edges and the related XMCD spectra for the sample of 10 uc LSMO/NNO recorded at 10K and $\mu_0H$=8T, (d-f) the element specific magnetic hysteresis loop, recorded as the field dependent XMCD intensity at Mn, Ni, and Nd edges (see red marks in a-c) for T=10 and 100K. Multiplet simulations of XAS and XMCD for Nd$^{3+}$ are also shown in (c) together  with the experimental spectra.}}
 \label{fig:3}
\end{figure}  

\section{Exchange coupling at the interface}
Having established the valence of the Mn$^{4+}$ and Ni$^{2+}$ at the interface of LSMO/NNO, as expected from the charge transfer scenario, we evaluate in this section the interfacial exchange coupling behavior by exploring the element specific XMCD spectra of Mn, Ni, as well as of the Nd ions. The normalized XAS of Mn L$_{2,3}$, Ni L$_{2,3}$, and Nd M$_{4,5}$ and the related XMCD spectra for the sample of (10 uc LSMO)/NNO recorded at 10K and $\mu_0H$=8~T are shown in Fig.~3a-3c, respectively. The negative XMCD sign at both the Mn and the Ni L$_3$ edges, indicates that the Ni$^{2+}$ and Mn$^{4+}$ ions are aligned ferromagnetically in agreement with  a positive sign of the  Mn$^{4+}$-O$^{2-}$-Ni$^{2+}$ superexchange interaction at the interface, similar to that observed in La$_2$NiMnO$_6$\cite{Dass2003, Das2008}. Noncollinear magnetic order of Nd$^{3+}$ in epitaxial NdNiO$_3$ film has been inferred from resonant soft x-ray diffraction~\cite{Scagnoli2008}. Here, the M$_{4,5}$ edge XAS and XMCD spectra have been measured to investigate the magnetic properties of the Nd ions. The M$_{4,5}$ edge XAS spectra (Fig.~3c) at the interface show the same multiplet structures as obtained through numerical  simulations for the  Nd$^{3+}$ \cite{Thole1985}. The positive and negative signal of the XMCD spectra at the M$_5$ and M$_4$ edges is due to the antiparallel alignment between the orbital and spin moments of Nd. The orbital magnetic moments, which are higher in absolute value as compared the spin moments, are pulled parallel to the external field.

XMCD sum-rules \cite{Thole1992, Carra1993} have been applied to both the Ni and Mn spectra. The orbital moments are close to zero for both elements, since they are known to be  quenched by the crystal field. For the analysis of the Ni moments we use a number of holes equal to $n_h$=2.2, which lies in between the value of 2.5 for Ni$^{3+}$ in nickelates~\cite{Scagnoli2006} and 1.8 for Ni$^{2+}$ \cite{Saitoh1995}. The total magnetic moment moment of Ni $\rm M=M_S+M_L$ is determined to be 0.45$\pm 0.10 \mu_B$/Ni at 10K and $\mu_0H=8T$, which is much higher as compared to the reported value for the LMO/LNO interface. Noticed that  the same correction factor of 1.1 was used correct for the mixing of the L$_3$ and L$_2$ edges of Ni~\cite{Piamonteze2015}. For Mn, we used $n_h$=6.0, which lies in between the value of 5.5 for Mn$^{3+}$ and 6.4 for Mn$^{4+}$~\cite{Saitoh1995}. The total moment of Mn $\rm M=M_S+M_L$ is determined to be 3.0$\pm 0.20\mu_B$/Mn at 10K and $\mu_0H=8T$, similar to the value of Mn moments in LMO/LNO interface when the same correction  factor of 1.7 is used as in Ref.~\cite{Piamonteze2015}.   

For Nd M$_{4,5}$ edges the sum-rules~\cite{Thole1992, Carra1993} cannot confidently be applied without considering additional correction factors~\cite{Teramura1996}. Therefore, atomic multiplet calculations (Fig.~3c) using the QUANTY code ~\cite{Haverkort2014, note1} have been performed to obtain the spin and orbital moments for Nd. The experimental XAS spectrum was well reproduced by the simulations when considering a 4f$^{3}$ (L=6, S=3/2, J=L-S=9/2)  ground  state  configuration. After scaling the measured XMCD spectrum (at 10~K) by a factor of four,  the simulated XMCD spectra overlaps very well with the experimental data. The full saturated expectation values for Nd$^{3+}$ ions are M$_L$ = 3.8, M$_S$ =-2.2 and M=1.6$\mu_B$/ atom. By matching the experimental and the simulated spectra we obtain a total magnetic moment M=0.52$\mu_B$/Nd with the orbital (spin) moment equal to 1.23 (-0.71)$\mu_B$ aligned parallel (antiparallel) to the field. Thess values suggest that Nd net moment cannot be fully saturated within the available magnetic fields due to its intrinsic antiferromagnetic ordering.

The element specific magnetic hysteresis loops, recorded as the field dependent XMCD intensity at Mn L$_3$, Ni L$_3$, and Nd M$_4$ edges (see red marks in Fig.~3a-c), are shown in Fig.~3d-f for T=10 and 100K. The similarity of the hysteresis loop shape measured for Mn and Ni demonstrates a strong ferromagnetic coupling between the Mn and the Ni cations at the interface for both T=10~K and 100~K temperatures. Because NNO is antiferromagnet at low temperatures, the ferromagnetic ordering of Ni cations as well as their high magnetic moment can only be located at the interface in these heterostructures. 

\begin{figure}[t]
\includegraphics[width=0.9\linewidth]{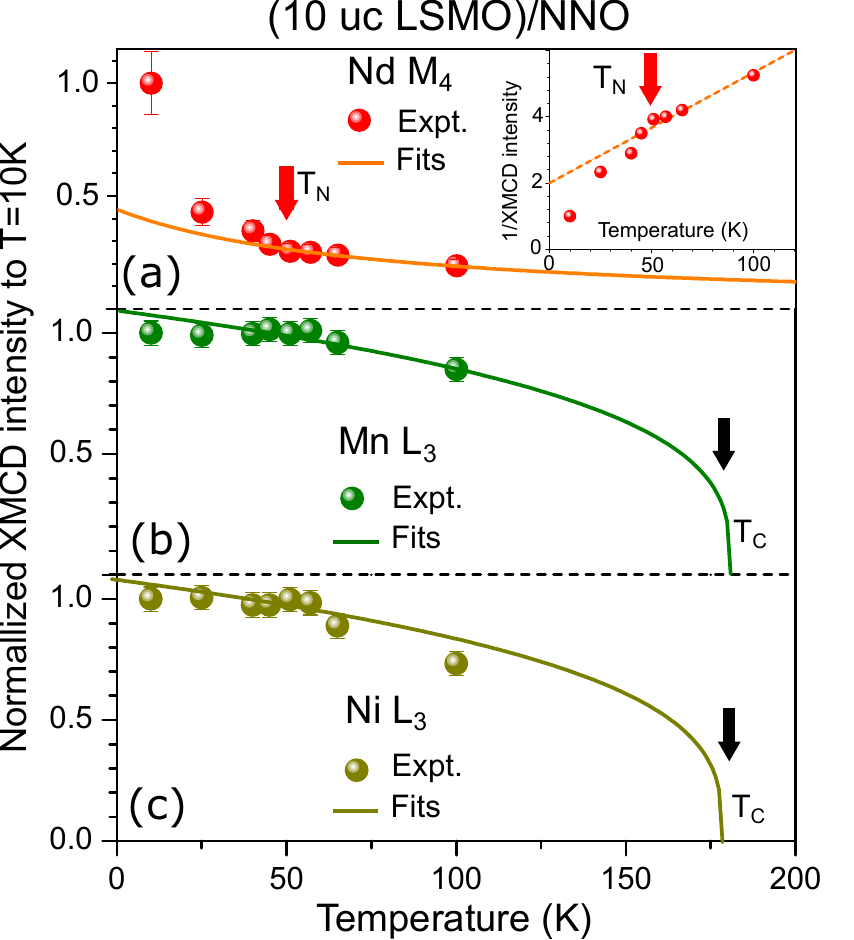}
\caption{\small{(a-c) The XMCD intensity measured at $\mu_0H$=8T for Nd M$_4$, Mn L$_3$, and Ni L$_3$ edges, varying the temperature from 10 to 100K. For Mn and Ni cations, the curves can be well fitted with $M\propto (T_c-T)^{\beta}$ with the exponent $\beta$=0.31 and $T_c=180 K$. For Nd ions, the canting moment is continuously increasing from 100 to 50K following the curve of $M\varpropto1/(T-\theta)$ with $\theta$=-75 K, while the inverse of the XMCD intensity (inset of Fig.~4a) deviates from the linear response at T=50 K which is the antiferromagnetic ordering temperature of the NNO films.}}    
\label{fig:4}
\end{figure}

On contrast, only linear magnetic response was observed for Nd cations for T=10 and 100K. The linear dependence of the element specific magnetization as a function of the magnetic field suggests either a paramagnetic or an antiferromagnetic behaviour for the Nd moments. This is different when compared to the ferromagnetic behavior of Ni and Mn cations at the interface. To distinguish between these two scenario, temperature dependence of the interfacial magnetic susceptibility needs to be performed, which will be described further below.

The XMCD intensities measured at $\mu_0H$=8T for Nd M$_4$, Mn L$_3$, and Ni L$_3$ edges as a function of  temperature ranging from 10 to 100K are shown in Fig.~4a-c. For Mn and Ni cations, the curves follow the VSM curve of (10uc LSMO)/NNO sample (see Fig.~1c),and can be well fitted with $M\propto (T_c-T)^{\beta}$ with the exponent $\beta$=0.31 and $T_c=180 K$. To establish the antiferromagnetic ordering of the NNO layer, we have measured the magnetization of the Nd element as a function of temperature, as shown in Fig.~4a.  We observe that the magnetization deviates significantly from an inverse linear behavior at 50 K. Plotting the inverse of the XMCD intensity as a function of temperature, shown in the inset of Fig.~4,  we observe a typical behavior of a magnetic susceptibility character, where deviations from a paramagnetic linear behavior is indicative for a paramagnetic to antiferromagnetic phase transition. As such, we are able to conclude that the onset of an antiferromagnetic ordering of the buried NNO films takes place and the N{\'e}el temperature of NNO which is determined to be 50~K. 

Note that the temperature dependence of the magnetization of the NNO layer, recognized also as  different slopes for the 10K and 100K magnetization curves shown in Fig.~3f, help to resolute on the nature of the linear increase of magnetization for the Ni and Mn hysteresis loops at high fields (see Fig.~3d and Fig.~3e). The slope of the hysteresis loop for Ni and Mn at high fields are not changing as a function of temperature. This suggests that coupling to the NNO layer is not likely to be the origin of this slope, otherwise it would change across the N{\'e}el temperature, similar to the slope change which is characteristic to the AF layer. This further indicates that most plausible origin of this slope can resides in antiferromagnetic correlations of Mn, belonging to the LSMO layer itself. They can be located at the interface due to a variation of the Mn valence, or even be part of the LSMO layer as a concurrent phase to the ferromagnetic ordering. Note that discussions on a similar effect, which was observed for the LSMO/LNO system, left the origin of this effect open~\cite{Sanchez2012,Piamonteze2015}.

The T$_{MIT}$ and T$_N$ were reported to be equal for  bulk NNO~\cite{Torrance1992} crystals, which are close 180~K. For thin films, both the T$_{MIT}$ and T$_{N}$ are reduced, with T$_{N}\leq$T$_{MIT}$\cite{Alsaqqa2017}. In our case, T$_N$=50~K is far below the T$_{MIT}$ of 120~K. Therefore, the exchange bias effect which was only observed below a blocking temperature of  30~K, is related to the paramagnetic-antiferromagnetic transition of the NNO layer. The Ni$^{2+}$ at the interface can rotate due to the exchange interaction with the Mn element which can further be reversed by an relatively low external magnetic field. Besides, the maximum of the exchange bias field for LSMO/NNO is around 120 Oe, suggests that the pinned spins responsible for the exchange bias are reduced at the interface. As reported recently, using different substrates, it is possible to tune the charge transfer and magnetism at the NNO/LSMO interface~\cite{Xu2018}. According to the charge transfer scenario, an electron is donated from Mn to Ni to form Mn$^{4+}$-O-Ni$^{2+}$. On one hand, one may observe a reduced magnetization, since the saturated magnetization of Mn$^{4+}$ (S=3/2) is weaker as compared to Mn$^{3+}$ (S=2). On the other hand, the ferromagnetic Ni$^{2+}$ with the moment of 0.45$\mu$B/atom is formed at the interface replacing some  of the antiferromagnetic Ni$^{3+}$, an therefore providing an even further reduction for the unidirectional anisotropy and favouring the  occurrence of loose interfacial spins responsible for an enhanced coercive field as shown in Fig.~1g.

\section{Conclusions}
In conclusion, using  soft x-ray spectroscopy and XMCD in high magnetic fields  we have shown that interfacial charge transfer from Mn$^{3+}$ in LSMO to Ni$^{2+}$ in NNO drives a ferromagnetic coupling Ni$^{2+}$-O$^{2-}$-Mn$^{4+}$ at the LSMO/NNO interface. Analysing the line shape changes of  the temperature dependent XAS spectra, the MIT was resolved and its  temperature onset was determined to be $T_{MIT}$=120K. The occurrence of antiferromagnetic phase of the NNO layer was established and characterised by element specific magnetic susceptibility measurements. The onset of the AF order was measured to occur at a the  N{\'e}el temperature equal to T$_N$=50K. A ferromagnetic coupling between the interfacial Ni$^{2+}$ and the Ni$^{3+}$ ions is observed, with the later being antiferromagnetically ordered in the deeper NNO monolayers. Our findings strongly suggest that the interfacial charge transfer plays an important role for the interfacial magnetism and can be used for tuning the magnetic properties of the upper ferromagnetic layers. The exchange bias effect observed in this system, below T=30K, is reduced due to the frustrated nature of the interface. 

We thank the HZB for the allocation of synchrotron radiation beamtime. The authors acknowledge the financial support for the VEKMAG project and for the PM2-VEKMAG beamline by the German Federal Ministry for Education and Research (BMBF 05K10PC2, 05K10WR1, 05K10KE1) and by HZB. Steffen Rudorff is acknowledged for technical support.


\begin{thebibliography}{99}
\bibitem{note1}\small{Parameters:  Oh symmetry, $\mu_0$H=6 T, T=0.1K, the scale factor for Slater integrals of Hartree-Fock values and the spin orbital coupling are 0.8 and 1, respectively.}
\end{thebibliography}
\end{document}